\documentclass[twoside]{iopart}
\usepackage[T1]{fontenc}
\usepackage[latin9]{inputenc}
\usepackage{geometry}
\usepackage{graphicx}

\makeatletter

\usepackage{iopams}
\usepackage{setstack}

\usepackage[numbers, sort&compress]{natbib}

\def\newblock{\hskip .11em plus .33em minus .07em}

\usepackage{bm}

\makeatother

\begin{document}

\title{Pinning and switching of magnetic moments in bilayer graphene}

\author{Eduardo V Castro$^{1,2}$,
M P L\'opez-Sancho$^{1}$, and M A H Vozmediano$^{1}$}

\address{{\large $^{1}$ }Instituto de Ciencia de Materiales de Madrid, CSIC,
Cantoblanco, E-28049 Madrid, Spain}

\address{$^{2}$ Centro de Física do Porto, Rua do Campo Alegre 687, P-4169-007
Porto, Portugal}

\ead{{ evcastro@icmm.csic.es, pilar@icmm.csic.es,
vozmediano@icmm.csic.es}}

\begin{abstract}
We examine the magnetic properties of the localized states induced
by lattice vacancies in bilayer  graphene with an unrestricted
Hartree-Fock calculation. 
We show that with realistic values of
the parameters and for experimentally accessible gate voltages we
can have a magnetic switching between an unpolarized and a fully
polarized system.

\end{abstract}




\maketitle

%

\section{Introduction}

\label{sec:Intro} After the explosion of publications on graphene
following the experimental synthesis \cite{Netal05,ZTSK05} the
present attention is centered on the experimental advances aiming
to generate better samples for electronic devices. One of the
major problems preventing applications of single layer graphene (SLG) is
the difficulty to open and control a gap in the samples. To this
respect bilayer graphene (BLG) and multilayer samples are more promising
\cite{Oetal08}. One of the potentially most interesting aspects of
graphene for the applications and that remains up to now partially
unexplored concerns the magnetic properties. Ferromagnetic order
enhanced by proton irradiation has been observed in graphite
samples \cite{Betal07} and demonstrated to be due to the carbon
atoms by dichroism experiments \cite{Oetal07}. By now it is clear
that the underlying mechanism leading to ferromagnetism in these
carbon structures is the existence of unpaired spins at defects
induced by a change in the coordination of the carbon atoms
(vacancies, edges or related defects) \cite{KM03}. Very recent
experiments on thin films in irradiated graphite show that the
main effect of proton irradiation is to produce vacancies on the
outer layers of the samples. For thin enough films of a few
thousands angstroms the protons go through the samples leaving
some vacancies behind. These samples show an enhanced local
ferromagnetism and also a better conductivity than the untreated
samples with less defects \cite{ASetal09}. Vacancies can play a
major role on these magnetic and transport properties and are
lately been recognized as one of the most important scattering
centers in SLG and BLG \cite{Metal09}.

The existence and nature of localized states arising from
vacancies in BLG have been analyzed in a recent paper
\cite{CLV09}. It was found that the two different types of
vacancies that can be present in the BLG system -- depending
on the sublattice they belong to --  give rise to two different
types of states: quasi-localized states, decaying as $1/r$ for
$r\to\infty$, similar to these found in the SLG case
\cite{PGS+06}, and truly delocalized states, going to a constant
as $r\to\infty$. When a gap is induced by the electric field
effect quasi-localized states give rise to resonances at the gap
edges while the delocalized ones become truly localized inside the
gap. These findings are very important in understanding the
magnetic properties of the graphitic samples since these localized
states carry magnetic moments. In this paper we study the magnetic
properties of the localized states found in \cite{CLV09} using an
unrestricted Hartree-Fock calculation. The most interesting case
arises in the presence of a gate (perpendicular electric field 
$\mathbf{E}=E_{z}\,\hat{e}_{z}$) 
opening a gap when considering two vacancies of the same sublattice 
located at different layers. We will show that with realistic values 
of the parameters and for experimentally accessible gate voltages we can
have a magnetic switching between unpolarized and fully polarized
system.

\section{The electronic  structure of bilayer graphene}

\label{sec:structure}
\begin{figure}
\begin{centering}
\includegraphics[width=0.85\columnwidth]{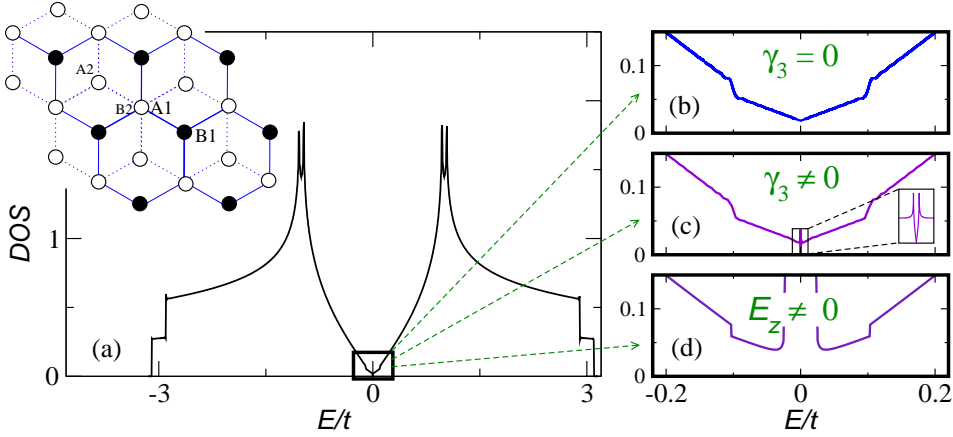}

\caption{\label{fig:bilayer}(a)~Total DOS of bilayer graphene. The
inset shows the lattice structure. (b)-(d)~DOS zoom at low
energies for the minimal model, for the model with $\gamma_{3}$,
and the model with finite gap due to perpendicular electric
field $E_z$, respectively. }

\par\end{centering}

\end{figure}

The lattice structure of a BLG is shown in
figure~\ref{fig:bilayer}. In this work we  consider only
$AB$-Bernal stacking, where the top layer has its $A$ sublattice
on top of sublattice $B$ of the bottom layer. We use indices~1
and~2 to label the top and bottom layer, respectively.

In the tight-binding approximation, the in-plane hopping energy, $t$,
and the inter-layer hopping energy, $\gamma_{1}$, define the most
relevant energy scales (see figure~\ref{fig:bilayer}). The simplest
tight-binging Hamiltonian describing non-interacting $\pi-$electrons
in BLG reads \cite{McClure57,SW58,MF06}:
\begin{equation}
H_{TB}=\sum_{i=1}^{2}H_{i}+\gamma_{1}\sum_{\mathbf{R},\sigma}
\big[a_{1,\sigma}^{\dagger}(\mathbf{R})b_{2,\sigma}(\mathbf{R})+
\mbox{h.c.}\big],
\label{eq:Hbilayer}
\end{equation}
with $H_i$ being the SLG Hamiltonian
\begin{equation}
H_{i}=-t\sum_{\mathbf{R},\sigma}\big[a_{i,\sigma}^{\dagger}(\mathbf{R})
b_{i,\sigma}(\mathbf{R})+a_{i,\sigma}^{\dagger}(\mathbf{R})
b_{i,\sigma}(\mathbf{R}-\mathbf{a}_{1})+
a_{i,\sigma}^{\dagger}(\mathbf{R})b_{i,\sigma}(\mathbf{R}-\mathbf{a}_{2})+
\mbox{h.c}\big],
\label{eq:Hslg}
\end{equation}
where $a_{i,\sigma}(\mathbf{R})$ {[}$b_{i,\sigma}(\mathbf{R})$] is
the annihilation operator for electrons at position $\mathbf{R}$
in sublattice $Ai$ ($Bi$), $i=1,2$, and spin $\sigma$. 
The basis vectors may be written as
$\mathbf{a}_{1}=a\,\rm{\hat e}_{x}$ and 
$\mathbf{a}_{2}=a(\rm{\hat e}_{x}-\sqrt{3}\,\rm{\hat e}_{y})/2$,
where $a=0.246\,\rm{nm}$.
The
estimated values of the  parameters for this minimal model are: 
$t\approx3\,\mbox{eV}$,
$\gamma_{1}\approx0.3\,\mbox{eV}\sim t/10$ \cite{NGPrmp}.

The main additional tight binding parameters are the inter-layer
second-nearest-neighbor hoppings $\gamma_{3}$ and $\gamma_{4}$
shown in figure~\ref{fig:bilayer} which play an important role in
what follows. $\gamma_{3}$ connects different sublattices
($B1-A2$) and $\gamma_{4}$ connects atoms of the same  sublattices
($A1-A2$ and $B1-B2$). Their values are less well known but  we
can assume  that the following relation between parameters holds,
$\gamma_{4}\sim\gamma_{3}\sim\gamma_{1}/3 \sim t/30$.

For the analysis of the bound states and associated magnetic
moments of the present work  a summary of the most relevant issues
of the BLG electronic structure is the following:

\begin{itemize}

\item The minimal model with only $\gamma_1$ has electron-hole
symmetry and it is a  bipartite lattice although not all the $A$ and
$B$ atoms are equivalent since some have (have not) a hopping to the
other layer. It  has two degenerate stable Fermi points similar to
SLG \cite{MGV07} but  the dispersion relation
around them is quadratic and the density of states (DOS) at the
Fermi points is finite (figure~\ref{fig:bilayer}(b)). 
Opening of a gap gives rise to the DOS
shown in figure~\ref{fig:bilayer}(d) with the characteristic double
minimum shape \cite{MAF07}. The value of $\gamma_1$ sets a bound
on the maximal value of the gap.

\item Inclusion of a $\gamma_3$ together with $\gamma_1$ coupling
lifts the degeneracy of the Fermi points that are shifted in
momentum space.  The dispersion relation around the Fermi points
is linear and the DOS is zero very much like in the SLG
case (figure~\ref{fig:bilayer}(c)). 
The lattice is still bipartite in the sense that,
generically, atoms of type $A$ are only linked to atoms of type $B$
although the layer index and couplings make some differences
between different $A$ ($B$) atoms.

\item The  combination $\gamma_1$-$\gamma_4$  breaks the bipartite
nature of the lattice. On the electronic point of view it induces
an electron-hole asymmetry but the DOS at the Fermi point does not
change. This coupling is important for the magnetism of the
samples.

\end{itemize}

\section{The model and the magnetic properties of the perfect lattice}
\label{sec:model}

In order to study the magnetic behaviour of BLG in the presence of
vacancies and/or topological defects we use the Hubbard model,
treated in the Hartree-Fock approximation. The total Hamiltonian
then reads $H=H_{TB}+H_{U}$, where the on-site Coulomb part is
given by
\begin{equation}
H_{U}=U\sum_{\bm R,\iota}[n_{a\iota\uparrow}(\bm R)n_{a1\iota\downarrow}
(\bm R)+n_{b\iota\uparrow}(\bm R)n_{b\iota\downarrow}(\bm R)]\,,
\label{eq:Hubb}
\end{equation}
where $n_{x\iota\sigma}(\bm R)=x_{\iota\sigma}^{\dag}(\bm
R)x_{\iota\sigma}(\bm R)$, with $x=a,b$, $\iota=1,2$ and
$\sigma=\uparrow,\downarrow$. We use finite clusters with periodic
boundary conditions at half-filling (one electron per atom).

It is well known that the Hartree-Fock-RPA approximation for SLG
 produces a phase transition at the critical Hubbard interaction
 $U_{c}\approx2.2t$, above which the staggered magnetization becomes
finite \cite{Fetal96,PAB04}.
\begin{figure}
\begin{centering}
\includegraphics[width=0.4\columnwidth]{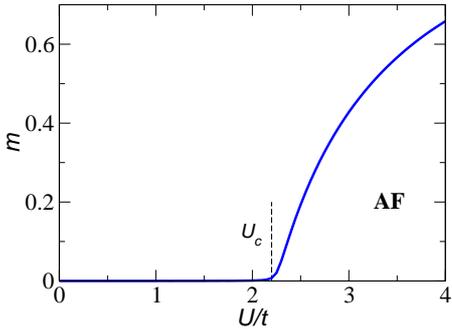}%
\begin{minipage}[b][1\totalheight]{0.6\columnwidth}%
\caption{\label{fig:mUbilHF}
Sublattice magnetization $m=|n_{\Gamma i,\uparrow}-n_{\Gamma i,\downarrow}|$
vs $U$ for bilayer graphene in the Hartree-Fock approximation, where
$\gamma_{1}=0.1t$ and $\Gamma i=A1,B1,A2,B2$. }
\end{minipage}%

\par\end{centering}

\end{figure}

The antiferromagnetic transition in the BLG case has been
analyzed in \cite{NNP+05}. Figure~\ref{fig:mUbilHF} shows the
sublattice magnetization as a function of the Hubbard repulsion
$U$. Throughout this work we will explore the magnetic behaviour of
the system with vacancies  for values of $U \leq t$ deep into the
region of $U$ where the sublattice magnetization is exponentially
suppressed so that we can attribute any magnetic moment to the
presence of defects.

\section{Vacancies in bilayer graphene}
\label{sec:vacancies}

Unlike the case of clean undoped SLG where the
 DOS at the Fermi level is zero and there is no
gap, in the BLG case and depending on the more relevant tight
binding parameters (see figure 1) we can have either a constant DOS
-- minimal model with only t and $\gamma_0$ -- or a zero 
DOS in the presence of $\gamma_3$. Moreover a gap can be easily
generated by an electric field perpendicular to the plane, as
mentioned before. The DOS
is crucial for the study of localized states. In the SLG
case, single vacancies  induce quasi-localized states around the
defect, decaying as $1/r$ \cite{VLSG05,PGS+06}. Due to the absence
of a gap, true bound states do not exist in the thermodynamic
limit.

In the BLG case there are two types of vacancies $beta$ and
$alpha$ for sites connected (or not) to the other layer. As shown
in \cite{CLV09} associated to the presence of vacancies
and to the existence of a gap in the spectrum generated by an
electric field $E_z$ three different types of vacancy-induced states are
found:
\begin{enumerate}
\item For $E_z=0$ a $\beta-$vacancy induces a resonance for $\gamma_3 =
0$ (delocalized state) and a quasi-localized state ($1/r$ behaviour)
for a finite value of $\gamma_3$.
\item For $E_z=0$ an $\alpha-$vacancy induces always a resonance
irrespective of $\gamma_3$.
\item For  $E_z\neq 0$ a $\beta-$vacancy produces a resonance inside the
continuum near the band edge while an $\alpha-$vacancy gives rise to a
truly localized state inside the gap. This is the most interesting
state for the magnetic implications.
\end{enumerate}

\begin{figure}
\begin{centering}
\includegraphics[width=0.99\columnwidth]{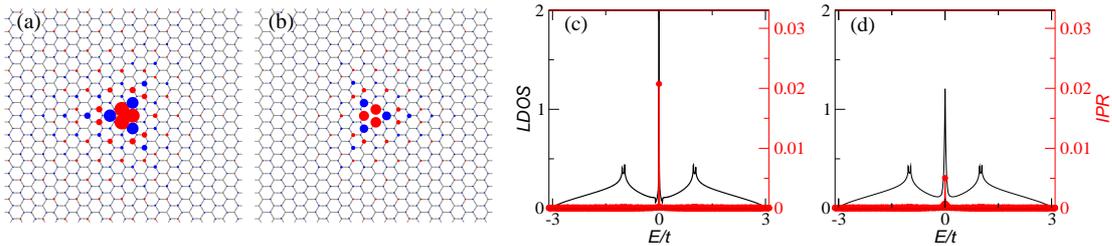}
\par\end{centering}

\caption{\label{fig:wfzm}(a)-(b) Zero-energy eigenstates in a
minimal model bilayer graphene cluster with $\gamma_{1}=0.1t$
containing a single $\beta-$ (a) and $\alpha-$vacancy (b). We show
only the  region around the vacancy and the layer where the
vacancy is located. 
(c)-(d) Local density of states (right axis) and inverse
partition ratio (left axis) for a cluster 
with a single $\beta-$ (c) and $\alpha-$vacancy (d) including
$\gamma_3 = 0.033t$.}

\end{figure}

In figure \ref{fig:wfzm}(a) and \ref{fig:wfzm}(b) 
we illustrate the nature of the
localized states 
by plotting
the numerical 
wavefunction for zero-energy
eigenstates in a BLG cluster
with a single $\beta-$ and $\alpha-$vacancy, respectively. 
The cluster contains $2\times74^2$ sites.
The image
shows only the layer where the vacancies are located and the
region around the vacancies. It can be seen that  a
quasi-localized state  exists for a $\beta-$vacancy, and for an
$\alpha-$vacancy the zero-energy mode also appears quasi-localized
over the diluted layer. The reduced amplitude of the zero mode in
figure~\ref{fig:wfzm}(b) for the $\alpha-$vacancy is due to the
presence of a delocalized component  on the underlaying layer (not
shown). 

The above results are confirmed by the enhanced local DOS
and enhanced inverse partition ratio at zero energy for a $\beta-$vacancy
(quasi-localized state),
as shown in figure~\ref{fig:wfzm}(c). The equivalent result for a
$\alpha-$vacancy (delocalized state with a quasi-localized component in
one of the layers) is shown in figure~\ref{fig:wfzm}(d). The local DOS
is computed at a site closest to the vacancy using the recursive
Green's function method for a cluster with $2\times1400^2$ sites.
The inverse participation ratio, defined as fourth moment
of the wavefunction amplitude, is computed for the cluster used
in figures~\ref{fig:wfzm}(a) and~\ref{fig:wfzm}(b).

\section{Magnetic behaviour}
\label{sec:magnetic}

The generation and structure of the magnetic moments associated to
unpaired atoms in SLG and multilayer graphene is to a great
extent determined by the bipartite nature of the underlying
lattice and hence by the Lieb's theorem \cite{L89}. The theorem
states that the ground state of the repulsive half filled Hubbard
model in any bipartite lattice with $N=N_{A}+N_{B}$ sites  is
unique and has total spin $S=\frac{1}{2}|N_{A}-N_{B}|$. According
to the Lieb's theorem \cite{L89} the quasi-localized zero modes
induced by unpaired atoms in the bipartite lattice become
spin-polarized in the presence of a Hubbard repulsion $U$ and
local moments appear in the lattice
\cite{MNFL04,LFetal04,VLSG05,PFB08,LJV09}. In the thermodynamic
limit the spin polarized modes (no longer at zero energy) merge
into the continuum and even though Lieb theorem applies equally,
the spin polarization is delocalized and itinerant ferromagnetism
appears \cite{L89}.

Pinning of magnetic moments in localized regions, in the
thermodynamic limit, would be a very  interesting possibility for
applications. In SLG we could try to open a gap
and push the quasi-localized modes out of the continuum. However,
a gap is not easily open in graphene, and a mass-gap does not
work: the same linear algebra theorem that guarantees the
existence of zero modes when no diagonal terms exist in the
Hamiltonian \cite{PdSN07} also states that, in the presence of a
staggered (diagonal) potential, these modes move to the gap edges;
this is due to the fact that these modes live only on one
sublattice, the less diluted. In BLG we can easily
open a gap by inducing layer asymmetry through the application 
of a perpendicular electric field $E_{z} \neq 0$ (back gate, 
for example), which is not a staggered potential. 
In \cite {CLV09} it was proven that truly localized
states exist inside the gap induced through electric field
effect in BLG.

\subsection{Single vacancy}

The magnetic properties of  a vacancy in BLG were
studied in \cite{CJKK08} using spin-polarized density functional
theory. It was found that the spin magnetic moment localized at
the vacancy is of the order of  ten percent smaller than  that of
SLG for both types of vacancies $\alpha$ and
$\beta$. This reduction of the spin magnetic moment in the bilayer
was attributed to the interlayer charge transfer from the adjacent
layer to the layer with the vacancy. We have verified that both in
the minimal model $\gamma_{3}=0$ and for a finite value of
$\gamma_{3}$ we obtain the results expected for a single
layer in accordance with Lieb theorem. The second-nearest-neighbor
hopping $\gamma_4$ that  breaks the bipartite character of the
whole lattice does not change this behaviour, as long as one vacancy
is considered. A finite $\gamma_4$ makes the vacancy-induced state
to appear off zero energy, but still spin degenerate. Including $U$ 
lifts spin degeneracy and induces a magnetic ground state. 
The situation is similar to the one discussed in 
SLG \cite{LJV09} when a pentagon is included.

\subsection{Two vacancies and the effect of an asymmetry gap}

Regarding the effect of two vacancies, we have found rather
different behaviour depending on whether an asymmetry gap is
present or not, and depending on the combination layer/sublattice
where the two vacancies occur. As mentioned before, such an
asymmetry gap is induced by making the two layers asymmetric, for
example, by applying a perpendicular electric field through a back
gate voltage. The resultant electrostatic energy difference
between layers $eE_zd$ ($d=0.34\,\rm{nm}$ is the interlayer distance
and $e$ the electron charge) introduces an interesting tuning capability
to the system since all other parameters, including the strength
of the Hubbard interaction $U$, can hardly be tuned in
experiments.

\begin{figure}
\begin{centering}
\includegraphics[width=0.5\columnwidth]{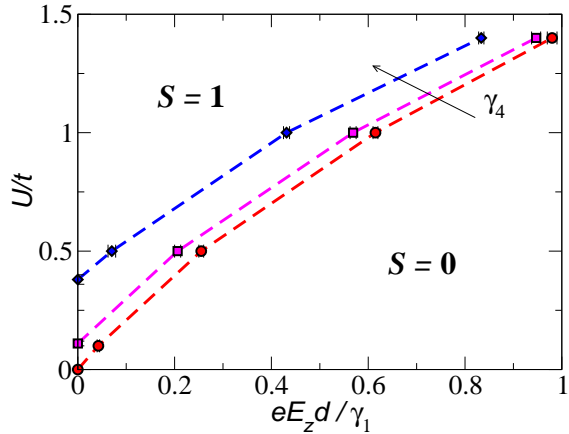}
\par
\end{centering}
\caption{\label{fig:VcU}Transition line between total spin $S=1$
and $S=0$ in the $E_{z}-U$ plane for the ground state of a bilayer
system with two vacancies belonging to the same sublattice located
in different layers for different values of the $\gamma_4$ hopping
integral. Circles (blue) , squares (purple), and diamonds (red)
stand for $\gamma_{4}=0,\gamma_{1}/4, \gamma_{1}$, respectively.
The error bar is of the order of symbol size.}
\end{figure}

For the bipartite case where $\gamma_{4}=0$ results are in
complete agreement with Lieb theorem, as expected. In particular,
irrespective of the layer index, two vacancies  of the same
sublattice induce a total spin $S=1$, while vacancies in different
sublattices give rise to a ground state with $S=0$. 
When
$E_z\neq0$, since there are nonzero diagonal elements in the
Hamiltonian matrix, we no longer have a bipartite system in the
Lieb sense \cite{L89} and different behaviour from these
determined by Lieb theorem might arise. We have found that two
vacancies in different sublattices always give $S=0$, irrespective
of the layer index and of the gap's size, in accordance with Lieb
theorem. Also following Lieb, vacancies in the same sublattice,
and belonging to the same layer, i.e. $Ai,Ai$ or $Bi,Bi$ with
$i=1,2$, originate $S=1$. 

The
interesting case arises when the system is gaped and has
two vacancies from the same sublattice in different layers,
i.e. $A1,A2$ or $B1,B2$. In this case we have two regimes: for
small $E_z$ we  get $S=1$ in agreement with Lieb theorem. When
$E_z$ increases we reach a regime at a critical value of  $E_z$
where the ground state has $S=0$.  Inclusion of the $\gamma_4$
hoping goes in the same direction as the gap depressing  the
polarization. The critical $U$ to maintain the full polarization of
the lattice increases for bigger values of $\gamma_4$. The
critical line in the $E_z-U$ plane is shown in figure~\ref{fig:VcU}
for different values of $\gamma_4$. The
explanation for this behaviour lies in the different ways in which the
degeneracy of the  zero modes is lifted with  $U$ -- that splits the
degeneracy according to the spin -- and with other couplings like $E_z$,
$\gamma_4$, or in-plane next-nearest-neighbor $t'$ \cite{KH07b}.
It is important to note that
the transition between finite and zero magnetic polarization of
the lattice occurs in the region of experimentally relevant values
of the external voltage: $0<E_z d\lesssim\gamma_{1}$ \cite{CNM+07}, 
which makes the
realization and observation of such magnetic switching capability
a real possibility.

\section{Conclusions and discussion}
\label{conclusions}

We have examined the magnetic properties of the localized states
induced by lattice vacancies in BLG recently analyzed
in \cite{CLV09}. We have found that in the presence of a gap
the system supports two types of spin polarized local states
related to the two types of inequivalent vacancies that can exist
in a Bernal stacking. Those living inside the gap are truly
normalizable bound states what can give rise to fully localized
large magnetic moments if there are several vacancies belonging to
the same graphene layer. This can be related to the measurement of
local magnetic moments in proton bombarded graphite associated to
the defects \cite{Eetal03b} and to the observation of the
insulating nature of the ferromagnetic regions \cite{Setal08}. A
density of such vacancies would give rise to a mid-gap band
contributing to the total conductivity of the sample. In such band
many body effects will be important and can drive the system to
other kinds of instabilities \cite{GKV08}. The other type of
vacancies stay at the edge of the gap and give rise to
quasi-localized magnetic moments whose wave function decays as $1/r$
similar to the states induced by vacancies in the monolayer
systems.

The most interesting case arises in the presence of a gate
opening a gap when considering two vacancies of the same
sublattice located at different layers. We have shown that with
realistic values of the parameters and for experimentally
accessible gate voltages we can have a magnetic switching between
unpolarized and fully polarized system.

Under the physical point of view our analysis can help to
understand the local ferromagnetism measured in thin films of
irradiated graphite \cite{Eetal03b} and the recent reports of an
increasing of the conductivity of the thin films of graphite after
irradiation with protons whose main effect is to produce vacancies
on the samples \cite{ASetal09}.


\ack{We thank F. Guinea for useful conversations. This research
has been partially supported by the Spanish MECD grant
FIS2005-05478-C02-01 and FIS2008-00124. EVC acknowledges financial
support from the Juan de la Cierva Program (MCI, Spain).}{}



\bibliographystyle{unsrt}

\addcontentsline{toc}{section}{\refname}
\begin{thebibliography}{10}

\bibitem{Netal05}
Novoselov K S, Geim A K, Morozov S V, Jiang D, Katsnelson M I, 
  Grigorieva I V, Dubonos S V and Firsov A A 2005
\newblock {\em Nature} {\bf 438} 197

\bibitem{ZTSK05}
Zhang Y, Tan Y W, Stormer H L and Kim P 2005
\newblock {\em Nature} {\bf 438} 201

\bibitem{Oetal08}
Oostinga J B, Heersche H B, Liu X, Morpurgo A F and Vandersypen L M K 2008
\newblock {\em Nat. Mater} {\bf 7} 151

\bibitem{Betal07}
Barzola-Quiquia J, Esquinazi P, Rothermel M, Spemann D, Butz T and
  Garc\'ia N 2007
\newblock {\em Phys. Rev.} B {\bf 76} 161403

\bibitem{Oetal07}
Ohldag H, Tyliszczak T, Höhne R, Spemann D, Esquinazi P, Ungureanu M and
  Butz T 2007
\newblock {\em Phys. Rev. Lett.} {\bf 98} 187204

\bibitem{KM03}
Kusakabe K and Maruyama M 2003
\newblock {\em Phys. Rev.} B {\bf 67} 092406

\bibitem{ASetal09}
Arndt A, Spoddig D, Esquinazi P, Barzola-Quiquia J and Butz T 2009
\newblock {\em Preprint} arXiv:0905.2954v1 [cond-mat.mtsl-sci]

\bibitem{Metal09}
Monteverde M, Ojeda-Aristizabal C, Weil R, Ferrier M, Guseron S,
Bouchiat H, Fuchs J N and Maslov D 2009
\newblock {\em Preprint} arXiv:0903.3285v2 [cond-mat.mes-hall]

\bibitem{CLV09}
Castro E V, L\'opez-Sancho M P and Vozmediano M A H 2009
\newblock {\em Preprint} arXiv:0906.4061v1 [cond-mat.mtrl-sci]

\bibitem{PGS+06}
Pereira V M, Guinea F, Lopes~dos Santos J M B, Peres N M R and
  Castro~Neto A H 2006
\newblock {\em Phys. Rev. Lett.} {\bf 96} 036801

\bibitem{McClure57}
McClure J W 1957
\newblock {\em Phys. Rev.} {\bf 108} 612

\bibitem{SW58}
Slonczewski J C and Weiss P R 1958
\newblock {\em Phys. Rev.} {\bf 109} 272

\bibitem{MF06}
McCann E and Fal'ko V 2006
\newblock {\em Phys. Rev. Lett.} {\bf 96} 086805

\bibitem{NGPrmp}
Castro Neto A H, Guinea F and Peres N R P 2009
\newblock {\em Rev. Mod. Phys.} {\bf 81} 109

\bibitem{MGV07}
Ma{\~n}es J L, Guinea F and Vozmediano M A H 2007
\newblock {\em Phys. Rev.} B {\bf 75} 155424

\bibitem{MAF07}
McCann E, Abergel D S L and Fal'ko V I 2007
\newblock {\em Solid State Comm.} {\bf 143} 110

\bibitem{Fetal96}
Fujita M, Wakabayashi K, Nakada K, Kusakabe K 1996
\newblock {\em J. Phys. Soc. Jpn.} {\bf 65} 1920

\bibitem{PAB04}
Peres N M R, Ara\'ujo M A N and Bozi D 2004
\newblock {\em Phys. Rev.} B {\bf 70} 195122

\bibitem{NNP+05}
Nilsson J, Castro~Neto A H, Peres N M R and Guinea F 2006
\newblock {\em Phys. Rev.} B {\bf 73} 214418

\bibitem{VLSG05}
Vozmediano M A H, L\'opez-Sancho M P, Stauber T and Guinea F 2005
\newblock {\em Phys. Rev.} B {\bf 72} 155121

\bibitem{L89}
Lieb E 1989
\newblock {\em Phys. Rev. Lett.} {\bf 62} 1201

\bibitem{MNFL04}
Ma Y, Lehtinen P O, Foster A S and Nieminen R M 2004
\newblock {\em New J. Phys.} {\bf 6} 68

\bibitem{LFetal04}
Lehtinen P O, Foster A S, Ma Y, Krasheninnikov A V and Nieminen R M 2004
\newblock {\em Phys. Rev. Lett.} {\bf 93} 187202

\bibitem{PFB08}
Palacios J , Fernandez-Rossier J and Brey L 2008
\newblock {\em Phys. Rev.} B {\bf 77} 195428

\bibitem{LJV09}
L\'opez-Sancho M P, de~Juan F and Vozmediano M A H 2009
\newblock {\em Phys. Rev.} B {\bf 79} 075413

\bibitem{PdSN07}
Pereira V M, Lopes~dos Santos J M B and Castro~Neto A H 2008
\newblock {\em Phys. Rev.} B {\bf 77} 115109

\bibitem{CJKK08}
Choi S, Jeong B W, Kim S and Kim G 2008
\newblock {\em J. Phys.: Condens. Matter} {\bf 20} 235220

\bibitem{KH07b}
Kumazaki H and Hirashima D S 2007
\newblock {\em J. Phys. Soc. Jpn.} {\bf 76} 064713

\bibitem{CNM+07}
Castro E V, Novoselov K S, Morozov S V, Peres N M R, Lopes dos Santos J M B,
Nilsson J, Guinea F, Geim A K and Castro~Neto A H 2007
\newblock {\em Phys. Rev. Lett.} {\bf 99} 216802

\bibitem{Eetal03b}
Esquinazi P, Spemann D, H\"ohne R, Setzer A, Han K H and Butz T 2003
\newblock {\em Phys. Rev. Lett.} {\bf 91} 227201

\bibitem{Setal08}
Schindler K, García N, Esquinaziand P and Ohldag H 2008
\newblock {\em Phys. Rev} B {\bf 78} 045433

\bibitem{GKV08}
Guinea F, Katsnelson M I and Vozmediano M A H 2008
\newblock {\em Phys. Rev.} B {\bf 77} 075422

\end{thebibliography}

\end{document}